\documentclass[prd,twocolumn,showpacs,preprintnumbers,amsmath,amssymb]{revtex4}

\usepackage{rotating}
\usepackage{graphicx}
\usepackage{dcolumn}
\usepackage{amstext}
\usepackage{lscape}
\usepackage{verbatim}
\usepackage[colorlinks,citecolor=blue,bookmarks]{hyperref}
\usepackage{times}
\usepackage{bm}
\usepackage{soul}

\begin{document}

\title{GW170817 constraints on the properties of a neutron star in the presence of WIMP dark matter}

\author{\large Abdul Quddus$^1$}
\author{\large Grigorios Panotopoulos$^2$}
\author{\large Bharat Kumar$^{3,4}$}
\author{\large Shakeb Ahmad$^1$}
\author{\large S.~K.~Patra$^{5,6}$}

\affiliation{$^1$Department of Physics, Aligarh Muslim University, Aligarh - 202002, India}
\affiliation{$^2$Centro de Astrof{\'i}sica e Gravita{\c c}{\~a}o, Departamento de F{\'i}sica, Instituto Superior T{\'e}cnico-IST,
Universidade de Lisboa-UL, Av. Rovisco Pais, 1049-001 Lisboa, Portugal}
\affiliation{$^3$Inter-University Centre for Astronomy and Astrophysics, Pune - 411007, India}
\affiliation{$^4$Center for Computational Sciences, University of Tsukuba, Tsukuba 305-8577, Japan}
\affiliation{$^5$Institute of Physics, Sachivalaya Marg, Bhubaneswar - 751005, India}
\affiliation{$^6$Homi Bhaba National Institute, Anushakti Nagar, Mumbai - 400085, India.}

\date{\today}

\begin{abstract}
The properties of a neutron star are studied in the presence of dark matter. We have considered a relatively 
light Weakly Interacting Massive Particle (WIMP) as a dark matter candidate with properties suggested by the results of the DAMA/LIBRA collaboration, realized for instance within the framework of the Next-to-Minimal Supersymmetric Standard Model. The dark matter particle interacts with the baryonic matter of a neutron star through Higgs bosons. The dark matter variables are essentially fixed using the results of the DAMA/LIBRA experiment, which are then used to build the Lagrangian density for the WIMP-nucleon interaction inside a neutron star. 
We have used the effective field theory motivated relativistic mean field model to study the equations-of-state in the presence of dark matter. The predicted 
equations-of-state are used in the Tolman-Oppenheimer-Volkoff equations to obtain the mass-radius relations, the moment of inertia, and effects of the tidal field on a neutron star. The calculated properties are compared with the corresponding data of the GW170817 event.  
\end{abstract}


\maketitle

\section{Introduction} 

One of the most chaotic and enthralling conundrum in physics is the problem of dark matter in the Universe. Several Cosmological and Astrophysical observations suggest that at least $90\%$ mass of the Universe is due to some non-luminous matter, yet to be discovered, the so-called “dark matter” (DM). The term ``dark matter" was coined by Zwicky in 1933 when he found some evidence about the missing mass in studying a cluster of galaxies 
known as ``Coma" \cite{zwicky}. He found that the expansion of the space (red-shift) 
in the Coma cluster could not be explained in terms of the known luminous mass. He applied the virial theorem and concluded that a large amount of DM must be present to keep these galaxies bound together. The measurements of the cosmic microwave background (CMB), too, suggest that DM is necessary to explain structure formation \cite{carlos}. Structure formation implies that clumps of neutral particles arose through 
the gravitational attraction, form neutral atoms which were attracted gravitationally by DM to form the galaxies. Another evidence of DM is the high temperature of the gas detected in clusters through its X-ray emission \cite{carlos,berg06}. Currently, there are a plethora of modern observations which support and confirm the 
existence of DM on a wide range of scales.
Many DM candidates have been proposed and studied over the years by cosmologists and particle physicists alike in an effort to constrain its properties. For a nice list of the existing DM candidates see e.g. \cite{taoso}, and for DM searches go through Refs. \cite{DM3,jennifer,DM5}. Despite that, the origin and nature of DM still 
remain unknown. Indeed, the determination of the type of elementary particles that play the role of DM in the Universe is one of the current challenges of Particle Physics and modern Cosmology.

WIMPs, which are thermal relics from the Big-Bang, are perhaps the most popular DM candidates. Initially, when the Universe was very hot, WIMPs were in thermal equilibrium with their surrounding particles. As the Universe expands and cools down, at a certain temperature which depends on the 
precise values of the mass of the DM particle and its couplings to the Standard Model (SM) particles, WIMPs decouple from the thermal bath, and its abundance freezes out. After freezing out, they can no longer annihilate, and their density is the same since then comprising the observed DM abundance of the Universe \cite{carlos}.
If WIMPs (let us have in mind the lightest neutralino in supersymmetric models, although the analysis and the obtained results still hold true for any fermionic WIMP) are the main candidates for DM, they will cluster gravitationally with stars, and also form a background density in the Universe. In Ref. \cite{smith90}, it was remarked that our own galaxy, the Milky Way, contains a large amount of DM. This raises the hope of 
detecting relic WIMPs directly, by scattering experiments in Earth-based detectors. The interaction of the DM particle with nuclei through elastic \cite{carlos} or inelastic scattering \cite{ellis88,stark95} is being studied in various 
laboratories. More than 20 experiments worldwide for DM direct detection searches are either running or in preparation, and some of them are the following: the DArk MAtter (DAMA) experiment \cite{dama1,dama2}, 
Cryogenic Dark Matter Search (CDMS) experiment \cite{cdms}, EDELWEISS experiment \cite{edel}, IGEX \cite{igex}, ZEPLIN \cite{zeplin}, 
 GErmanium DEtectors in ONe cryostat (GEDEON) \cite{morales}, CRESST \cite{cresst}, GErmanium in liquid NItrogen Underground Setup 
(GENIUS) \cite{genius}, and LHC. Furthermore, Fermi-LAT, GAMMA-400, IceCube, Kamiokande, and AMS-02 are some of the indirect DM detection experiments \cite{jennifer}.

Despite the null results of other experiments \cite{akerib}, which only put an upper limit on the nucleon-DM particle, the DAMA/LIBRA collaboration, located underground at the Laboratori Nazionali del Gran Sasso in Italy, has been reporting for many years an annual modulation caused by the variation of the velocity of the detector relative to the galactic DM halo as the Earth orbits the Sun. In particular, the final model-independent results of phase 1 were published in 2013 \cite{phase1}, while last year they published the first model-independent results of phase 2 \cite{phase2} collecting data from 6 annual cycles. In the second upgraded phase, which started at the end of 2010, the two main improvements in comparison with the first phase are the doubled exposure as well as the lower energy threshold from $2~keV$ to $1~keV$. If the signal reported by the DAMA/LIBRA collaboration is interpreted as WIMP DM, it gives rise to the $\sim 10~GeV$ WIMP hypothesis with a spin-independent (SI) DM-nucleon at $\sim 10^{-40}~cm^2$, see Fig.~1 of \cite{theor1,theor2}. To be more precise
\begin{eqnarray}
3 \times 10^{-41} cm^{2} \lesssim \sigma_p^{SI} \lesssim 5 \times 10^{-39} cm^{2}
\label{crossrange}
\end{eqnarray}
while the range of the mass of the DM particle is
\begin{eqnarray}
3 GeV \lesssim m_{DM} \lesssim 8 GeV .
\label{mdm}
\end{eqnarray}
Various studies have taken these results into account \cite{foot,feng,bott,avigno}. 
In the WIMP scenario, a one-to-one relation is seen between the SI direct detection rate and DM relic density if its elastic scattering on nuclei occurs dominantly through Higgs exchange \cite{sarah}.
The SI direct detection cross-section of elastic scattering of DM ($\Psi$) with nuclei is given by \cite{sarah}
\begin{eqnarray}
\sigma(\Psi N \rightarrow \Psi N) = \frac{y^2}{\pi} \frac{\mu^2_r}{v^2 M_h^4} f^2 m_N^2 , 
\label{sicross}
\end{eqnarray}
where $v=246~GeV$ is the Higgs vacuum expectation value, $m_N,M_h$ are the nucleon mass and the Higgs mass, respectively. The variables $y$, $f m_N/v$ are the Yukawa couplings for DM interaction with Higgs boson and the interaction of Higgs particle with 
the nucleons, respectively. Finally, $\mu_r= \frac{m_N m_{DM}}{m_N+m_{DM}}$ is the reduced mass of a nucleon-DM particle system. The unknown parameters entering Eq. \ref{sicross} may be fixed as follows: First, the mass of the DM particle as well as the DM-nucleon system are taken to be the ones suggested by the aforementioned results of the DAMA/LIBRA collaboration. Then, for a given mass of the Higgs boson the Yukawa coupling $y$ can be determined.

Neutron stars (NSs) are considered to be unique cosmic laboratories to explore 
properties of ultra-dense matter under extreme conditions of density and neutron-proton asymmetry. It is a well-known fact that the mass of a NS is dominated by the core contribution.
The properties of a NS are predicted by its equation-of-state (EOS), a certain relation between
energy density and pressure. EOSs are used in the Tolman-Oppenheimer-Volkoff (TOV) equations \cite{tov} to study the mass-radius relation and other physical quantities of NS. On the other hand, by measuring the mass and radius of NS (pulsar), its EOS can be constrained. Moreover, the tidal deformability $\Lambda$ of an NS from the GW170817 data \cite{abbott17}, the historical first detection of gravitational waves from the binary neutron-star (BNS) 
merger by the LIGO-Virgo collaboration, provides a new probe to the interior of NS and their nuclear EOS \cite{annala}.
Currently, the most accurate constraint on the high-density behavior of an EOS comes from the observations of a 
few massive pulsars (or GW170817) with maximum mass $M_{max}=(2.01 \pm 0.04)M_\odot$ (or $M_{max} \lesssim 2.17M_\odot$) 
\cite{marg17,antoni13}, respectively. However, Ref. \cite{rezzo18} has inferred the maximum mass $M \lesssim 2.16^{+0.16}_{-0.15} M_\odot$ 
using the quasi-universal relation between the maximum mass and the mass-shedding limit.
%

The consequences of DM inside NS have been discussed in the literature \cite{raj,gould,goldman,kouvaris08,kouvaris10,lava10,guver, 
ellis18,ellisplb}. These discussions include the effect of charged massive DM particle on NSs \cite{gould}, 
trapped WIMPs inside NSs \cite{goldman}, DM annihilation and its effect on NSs \cite{kouvaris08,kouvaris10,lava10}, 
or the collapse of an NS due to accretion of non-annihilating DM \cite{guver} etc. 
The possible effects of DM cores on certain properties of NS have been studied in \cite{ellisplb,ellis18} assuming 
different nuclear EOSs {as well as} different fractions of DM. In particular, first in \cite{ellisplb} using a mechanical 
model the authors found that the DM cores may produce a supplementary peak in the characteristic gravitational 
wave spectrum of neutron star mergers, and then in \cite{ellis18} they investigated the impact of Fermionic asymmetric DM 
as well as bosonic self-interacting DM on mass-to-radius profile, maximum mass and tidal deformability $\Lambda$ of NSs. 
In \cite{nelson}, the authors studied the properties of NS considering the DM halo and constrained the DM parameters using the GW170817 data. 
It was pointed out in \cite{panta} that the mass-radius relation of a NS can be affected in the presence of DM inside the object. The authors of \cite{panta,tuhin} considered, respectively, the Walecka model \cite{walecka74} and the NL3 \cite{lala97} EOSs, within the framework of relativistic mean field (RMF) theory, 
for the nucleonic part and fermionic DM inside the neutron star with additional self-coupling of Standard Model Higgs boson in Ref. \cite{panta}. 

In this paper, we have investigated for the first time the effects of DM inside a NS adopting the $\sim 10~GeV$ WIMP hypothesis as suggested by the results of the DAMA/LIBRA collaboration, which can be realized e.g. in the framework of the 
Next-to-Minimal Supersymmetric Standard Model (NMSSM). 
The EOSs of nuclear matter can be generated within the framework of an effective theory with 
appropriate degrees of freedom. In particular, for example, perturbative QCD \cite{aleksi,tyler}, non-relativistic Skyrme type 
density functional theory \cite{tond}, and relativistic mean field approach \cite{walecka74} are used to predict EOSs of nuclear matter.  
We have considered the effective-field theory motivated relativistic mean field
model (E-RMF) to generate the EOS of NS by considering the IOPB-I \cite{iopb1}, G3 \cite{G3}, and NL3 \cite{lala97} parameter sets. 
Recently as a relativistic microscopic approaches, E-RMF is used widely to predict the properties of finite nuclei as well as nuclear matter.
Here, in addition to the mass-radius relations, and the tidal deformability, we have analyzed the effects of DM on the moment 
of inertia of an NS. 
Our work differs from other similar works in two respects, namely i) we have considered more EOSs (especially G3 within E-RMF) for hadronic matter, and ii) we have studied the impact of DM on more NS observables.

The paper is organized as follows: The formalism adopted in this work is presented in Sec. \ref{theory}. 
In Sub-section \ref{RMF} we briefly present the E-RMF formalism with the EOS of nuclear matter for the parameters chosen.  
Sub-section \ref{EOS-NS} contains the full Lagrangian density with the inclusion of DM in NS, while in Sub-section \ref{tov-tidal}
the TOV equations together with the observables studied here are presented. Our numerical results are discussed in Sec. 
\ref{result}. Finally, we summarize and conclude our work in Sec. \ref{summary}.

\section{Formalism}{\label{theory}}

\subsection {Effective field theory motivated relativistic mean field model (E-RMF)}{\label {RMF}}

The relativistic mean field (RMF) theory \cite{serot97,gambhir90,ring96} is one of the widely used microscopic approaches 
to investigate the properties of a nuclear system, i.e., from finite nuclei to nuclear matter. In the RMF model nucleons 
interact through the exchange of mesons, which modify the nucleons' bare properties. Thus they are quasi-particles with 
medium-dependent effective mass and baryon chemical potential.  
The advantage of RMF over its non-relativistic counterpart is that it accounts for spin-orbit interaction automatically. 
The E-RMF model is nothing but the extension of simple RMF in which all possible types of mesons and their self and 
cross-coupling are considered \cite{furnstahl97,G3}. 
In this work we have used the E-RMF Lagrangian, including the contributions from $\sigma -$, $\omega -$ mesons up to 4$^{th}$ 
order of expansion, and $\rho -$ and $\delta -$ mesons with powers up to 2$^{nd}$ order, which has been shown to be a 
good approximation to predict observables of nuclear systems in a  considerably satisfactory way
\cite{furnstahl97,G3}.  
Here for the sake of completeness, we present the energy density and pressure for an infinite nuclear matter (INM) within E-RMF 
by applying the mean field approximations and curtailing the terms irrelevant to the nuclear matter. 
The energy density for INM within E-RMF is:
\begin{widetext}
\begin{eqnarray}
{\cal{E}}_{had.} & = &  \frac{2}{(2\pi)^{3}}\int d^{3}k E_{i}^\ast (k)+\rho_b  W+
\frac{ m_{s}^2\Phi^{2}}{g_{s}^2}\Bigg(\frac{1}{2}+\frac{\kappa_{3}}{3!}
\frac{\Phi }{M} + \frac{\kappa_4}{4!}\frac{\Phi^2}{M^2}\Bigg)
\nonumber\\
&&
 -\frac{1}{2}m_{\omega}^2\frac{W^{2}}{g_{\omega}^2}\Bigg(1+\eta_{1}\frac{\Phi}{M}+\frac{\eta_{2}}{2}\frac{\Phi ^2}{M^2}\Bigg)-\frac{1}{4!}\frac{\zeta_{0}W^{4}}
        {g_{\omega}^2}+\frac{1}{2}\rho_{3} R
 \nonumber\\
 &&
-\frac{1}{2}\Bigg(1+\frac{\eta_{\rho}\Phi}{M}\Bigg)\frac{m_{\rho}^2}{g_{\rho}^2}R^{2}-\Lambda_{\omega}  (R^{2}\times W^{2})
+\frac{1}{2}\frac{m_{\delta}^2}{g_{\delta}^{2}}\left(D^{2} \right),
        \label{eq20}
\end{eqnarray}
\end{widetext}
while the pressure for INM is given by,
\begin{widetext}
\begin{eqnarray}
P_{had.} & = &  \frac{2}{3 (2\pi)^{3}}\int d^{3}k \frac{k^2}{E_{i}^\ast (k)}-
\frac{ m_{s}^2\Phi^{2}}{g_{s}^2}\Bigg(\frac{1}{2}+\frac{\kappa_{3}}{3!}
\frac{\Phi }{M}+ \frac{\kappa_4}{4!}\frac{\Phi^2}{M^2}  \Bigg)
\nonumber\\
& &
 +\frac{1}{2}m_{\omega}^2\frac{W^{2}}{g_{\omega}^2}\Bigg(1+\eta_{1}\frac{\Phi}{M}+\frac{\eta_{2}}{2}\frac{\Phi ^2}{M^2}\Bigg)+\frac{1}{4!}\frac{\zeta_{0}W^{4}}{g_{\omega}^2}
  \nonumber\\
& &
+\frac{1}{2}\Bigg(1+\frac{\eta_{\rho}\Phi}{M}\Bigg)\frac{m_{\rho}^2}{g_{\rho}^2}R^{2}+\Lambda_{\omega} (R^{2}\times W^{2})
-\frac{1}{2}\frac{m_{\delta}^2}{g_{\delta}^{2}}\left(D^{2}\right),
        \label{eq21}
\end{eqnarray}
\end{widetext}
where $\Phi$, $D$, $W$, and $R$ are the redefined fields for $\sigma$, $\delta$, $\omega$, and $\rho$ mesons as
$\Phi = g_s\sigma $, $D=g_\delta \delta$, $W = g_\omega \omega^0$, and $R = g_\rho \vec{\rho}^0$ , respectively, while $E_{i}^\ast(k)$=$\sqrt {k^2+{M_{i}^\ast}^2} \qquad  (i= p,n)$
is the energy with effective mass ${M_{i}^\ast}^2=k_F^2 + M_i^2$, and $k$ is the momentum of the nucleon. 
The quantities $\rho_b$ and $\rho_3$ in Eq. \ref{eq20} are the baryonic and iso-scalar densities as defined in \cite{iopb1}.

\subsection{Equation of state for neutron star in the presence of dark matter}{\label{EOS-NS}}

The Lagrangian density for DM-nucleon interaction through the exchange of Higgs bosons $h$ is given by \cite{tuhin}
\begin{eqnarray}
{\cal{L}} & = & {\cal{L}}_{had.} + \bar \chi \left[ i \gamma^\mu \partial_\mu - M_\chi + y h \right] \chi + 
              \frac{1}{2}\partial_\mu h \partial^\mu h  
\nonumber\\
& &
- \frac{1}{2} M_h^2 h^2 + f \frac{M_n}{v} \bar \varphi h \varphi , 
\label{lag-tot}
\end{eqnarray}
where ${\cal{L}}_{had.}$ is the Lagrangian density for pure hadronic matter. The wave functions $\chi$ and $\varphi$ correspond to 
DM particle and nucleon, respectively. We have not considered the higher order terms of the Higgs scalar potential ($i.e.,$ $h^3$ and $h^4$), since in the mean field theory approximation these terms are negligible \cite{panta}.

The factor $f$ parameterizes the Higgs-nucleon coupling, and a complete expression for $f$ can be found in \cite{singlet3}. 
Following the lattice computations \cite{lattice1,lattice2,lattice3,lattice4}, we shall consider the central value $f=0.3$ in agreement with \cite{singlet3}. For the DM sector, we shall assume a mass range and an SI DM-nucleon cross-section suggested by the DAMA results \cite{dama1,dama2}. It is easy to verify that if we take the Higgs mass to be 125 GeV, the Yukawa coupling $y$ computed using Eq. \ref{sicross} lies in the non-perturbative regime. Therefore we have to assume a light Higgs boson with a mass $M_h=40~GeV$, so that $y < 1$. 
The authors of \cite{john10,basic1} have shown that such a scenario can be realized in the framework of the NMSSM in agreement with the rest of the experimental constraints.

Before we continue with our discussion, perhaps it should be useful to briefly mention here the basic features of the NMSSM \cite{nmssm1,nmssm2,nmssm3} (for a review see \cite{nmssm4}). It is a simple extension of the MSSM in which a singlet supermultiplet is added, and it is characterized by the following properties: i) It preserves the nice properties of the MSSM, i.e. it solves the hierarchy problem while at the same time 
it provides us with an excellent DM candidate, ii) it solves the $\mu$ problem \cite{muproblem}, and iii) there is a rich Higgs sector with 2 Higgs bosons more in comparison with MSSM. In particular, if some of the Higgs bosons have a significant singlino component they can be light, $M_H \leq 70~GeV$, without any contradiction to current experimental constraints \cite{teixeira}. As a matter of fact, it has been shown that in NMSSM it is possible to obtain a DM-nucleon SI cross-section as high as the one indicated by the DAMA/LIBRA results, precisely due to the exchange of light Higgs bosons, which cannot be achieved in the MSSM \cite{john10,basic1}.

Moreover, one may briefly summarize the current status of DM in SUSY models taking into account LHC searches as follows: Supersymmetric models have been under siege after the Higgs boson discovery \cite{higgs1,higgs2} combined with the lack of any signal for sparticles \cite{null1,null2}, pushing the SUSY spectrum in the multi-TeV region \cite{siege1,siege2}. In natural SUSY with low fine-tuning electroweak symmetry breaking, the lightest neutralino is higgsino-like with a mass at (100-300)~GeV \cite{baer1}, which has been excluded as a single
DM candidate \cite{baer2}, as its abundance is lower than the WMAP/PLANCK measured
value by a factor of 10-15 \cite{baer3}. A mixed axion/higgsino dark
matter scenario has emerged, in which the axion is the dominant DM component in the bulk
of the parameter space \cite{baer4}. On the other hand, in the framework of the NMSSM a
light singlino with a mass lower than $60~GeV$ is still a viable DM candidate in a few
regions of the allowed parameter space \cite{Ref1}, while the lightest CP-even Higgs
boson, which is predominantly a singlet, can be as light as $48~GeV$ \cite{Ref2}.

Solving the full Lagrangian density (Eq. \ref{lag-tot}) by the variational principle, and taking care of all the mean field and INM approximations \cite{tuhin,iopb1}, the energy density and pressure are given by
\begin{eqnarray}
{\cal{E}} & = &  {\cal{E}}_{had.} + \frac{2}{(2\pi)^{3}}\int_0^{k_F^{DM}} d^{3}k \sqrt{k^2 + (M_\chi^\star)^2 } 
\nonumber\\
& &
+ \frac{1}{2}M_h^2 h_0^2 .
\label{etot}
\end{eqnarray}
\begin{eqnarray}
P & = &  P_{had.} + \frac{2}{3(2\pi)^{3}}\int_0^{k_F^{DM}} \frac{d^{3}k k^2} {\sqrt{k^2 + (M_\chi^\star)^2}} 
\nonumber\\
& &
- \frac{1}{2}M_h^2 h_0^2 .
\label{ptot}
\end{eqnarray} 
The Fermi momentum of DM particles ($k_f^{DM}$) is taken to be constant throughout the calculation with the value fixed at 0.06 GeV, although in Refs. \cite{tuhin,panta} the authors have considered values within a certain range. 
The effective mass of the nucleon ($M^\ast$) is modified due to the interaction with the Higgs boson. The new effective mass of the nucleon $M^\star$ and the effective mass of the DM particle $M_\chi^\star$ are given by,
\begin{eqnarray}
M_i^\star &=& M_i + g_\sigma \sigma -\tau_3 g_\delta \delta - \frac{f M_n}{v}h_0, 
\nonumber\\
M_\chi^\star &=& M_\chi -y h_0.
\label{effmass}
\end{eqnarray}

So far the discussion on the EOS has been for INM. Now we present in the discussion to follow how to obtain the EOS for NS. 
In a neutron star, the Fermi momentum of neutrons and protons are different due to the different number densities of these particles. 
For the stability of NSs, the $\beta -$ equilibrium condition is imposed, which is given by,
\begin{eqnarray}
\mu_n &=& \mu_p +\mu_e,   \nonumber \\
\mu_e &=& \mu_\mu.
\label{bequi}
\end{eqnarray}
where, $\mu_n$, $\mu_p$, $\mu_e$, and $\mu_\mu$ are the chemical potentials of neutrons, protons, electrons, and muons, respectively. The muon comes into play when the chemical potential of the electrons reaches the muon rest mass 
and maintains the charge of NS as follows
\begin{eqnarray}
\rho_p = \rho_e +\rho_\mu. 
\label{charge}
\end{eqnarray}
The chemical potentials $\mu_n$, $\mu_p$, $\mu_e$, and $\mu_\mu$ are given by, 
\begin{eqnarray}
\mu_n &=& g_\omega \omega_0 + g_\rho \rho_0+ \sqrt{k_n^2+ (M_n^\star)^2},
\label{mun}
\end{eqnarray} 
\begin{eqnarray}
\mu_p &=&  g_\omega \omega_0 - g_\rho \rho_0+ \sqrt{k_p^2+ (M_p^\star)^2},
\label{mup}
\end{eqnarray}
\begin{eqnarray}
\mu_e &=& \sqrt{k_e^2+ m_e^2},
\label{mue}
\end{eqnarray}
\begin{eqnarray}
\mu_\mu &=& \sqrt{k_\mu^2+ m_\mu^2}.
\label{mumu}
\end{eqnarray}

The particle fraction is determined by the self-consistent solution of Eq. \ref{bequi} and Eq. \ref{charge}
for a given baryon density. The total energy density and pressure for $\beta-$ stable NS are given by, 
\begin{eqnarray}
{\cal {E}}_{NS} &=& {\cal {E}} + {\cal {E}}_{l} \nonumber \\
P_{NS} &=& P + P_l .
\label{eos-nstar}
\end{eqnarray}
Where, 
\begin{eqnarray}
{\cal {E}}_{l} &=& \sum_{l=e,\mu}\frac{2}{(2\pi)^{3}}\int_0^{k_l} d^{3}k \sqrt{k^2 + m_l^2 },
\label{elep}
\end{eqnarray}
and
\begin{eqnarray}
P_{l} &=& \sum_{l=e,\mu}\frac{2}{3(2\pi)^{3}}\int_0^{k_l} \frac{d^{3}k k^2} {\sqrt{k^2 + m_l^2}}
\label{plep}
\end{eqnarray}
are the energy density and pressure for leptons ($e$ and $\mu$).
The EOSs of NS (Eq. \ref{eos-nstar}) are used as the input (with the representation ${\cal {E}}_{NS}\equiv {\cal {E}}$ and $P_{NS}\equiv p$) 
to TOV equations to find the NS observables.  
 
\subsection{Mass, radius, tidal deformability, and moment of inertia of NS}{\label {tov-tidal}}

The structural properties of NS, such as the mass-to-radius profile, the tidal deformability, the moment of inertia are studied in this work. 
Given an EOS it is straightforward to calculate the mass and radius of the NS by using the TOV equations \cite{tov}. 
For slowly rotating objects, we make as usual for the line element the following ansatz \cite{molnvik}
\begin{eqnarray}
ds^2 &=& - e^{\nu(r)} dt^2 + e^{\lambda(r)} dr^2 + r^2 (d\theta^2+\sin^2\theta d\phi^2)
\nonumber \\
&&-2 \omega(r) (r sin \theta)^2 dt d \phi 
\label{metric}
\end{eqnarray}  
The TOV equations are given by,
\begin{equation}
e^{\lambda(r)} = \left(1- \frac{2 m}{r}\right)^{-1},
\label{lambda}
\end{equation}
\begin{equation}
\frac{d\nu}{dr}= 2 \: \frac{m + 4 \pi p r^3}{r (r-2 m)}, 
\label{nu}
\end{equation}  
\begin{equation}
\frac{dp}{dr} = - \frac{({\cal{E}}+p)(m+4\pi r^3p)}{r (r-2m)}, 
\label{pres}
\end{equation}
\begin{equation}
\frac{dm}{dr} = 4 \pi r^2 {\cal{E}}. 
\label{mass}
\end{equation}

The moment of inertia (MI) of NSs is obtained by solving the TOV equations along with the equation including the rotational frequency (given below). For a slowly rotating NS the MI is given by \cite{hartle67,lattimer01},
\begin{eqnarray}
I=\frac{8\pi}{3} \int_0^R r^4 ({\cal {E}}+p) e^{(\lambda-\nu/2)} \frac {\bar \omega}{\Omega} dr, 
\label {mi}
\end{eqnarray}
where $\Omega$ and $\bar \omega(r) \equiv \Omega - \omega(r)$ are the angular velocity and the rotational drag function, respectively, for a
uniformly rotating NS. The rotational drag function $\bar \omega$ meets the boundary condition,
\begin{equation}
\bar \omega (r=R)=1-\frac{2I}{R^3}, \; \; \; \; \; \frac{d\bar\omega}{dr}|_{r=0}=0 
\label{bound.}
\end{equation}

The quantity $\frac{\bar \omega}{\Omega}$, evolve in Eq. \ref{mi}, is the dimensionless frequency satisfying the equation
\begin{eqnarray}
\frac{d}{dr} \left(r^4 j \frac{d\bar \omega}{dr}\right) = -4 r^3 \bar \omega \frac{dj}{dr}, 
\label{unitlessw.}
\end{eqnarray}
with $ j= e^{-(\lambda + \nu )/2}$.

The tidal deformability of an NS is one of the most important measurable physical quantities. 
It characterizes the degree of deformation of NS due to the tidal field of its companion in BNS. 
During the last stage of NS binary, each component star of binary system develops a mass quadrupole due to  
the tidal gravitational field of the partner NS. 
The tidal deformability for $l=2$ quadrupolar perturbations is defined to be,
\begin{eqnarray}
\lambda_2 = \frac{2}{3}k_2 R^5 ,
\label{lam2}
\end{eqnarray}
where $R$ is the NS radius, and $k_2$ is the tidal love number which depends on stellar structure.  
The $k_2$ is calculated using the expression \cite{tanja},
\begin{eqnarray}
k_2 = \frac{8C^5}{5}(1-2C)^2 (2(1-C)+(2C-1)y_R) \times \nonumber \\
 \{ 4C^3 \left(13-11y_R+2C^2(1+y_R)+C(-2+3y_R) \right) \nonumber \\
 + 2C \left( 6-3y_R+3C (5y_R-8) \right) +3 (1-2C)^2 \nonumber \\
 \times (2+2C(y_R-1)-y_R)\log (1-2C) \}^{-1}, 
\label{k2}
\end{eqnarray}
where $C=M/R$ is the compactness of the NS, and $y_R=y(R)$ is obtained by solving the following 
differential equation
\begin{eqnarray}
r \frac{dy}{dr} + y^2 + y F(r) + r^2 Q(r)=0
\label{yeqn}
\end{eqnarray}
with 
\begin{eqnarray}
F(r) &=& \frac{r-4\pi r^3 ({\cal{E}}-p)}{r-2M},  \nonumber \\
Q(r) &=& \frac{4\pi r \left(5 {\cal {E}}+9 p + \frac{{\cal{E}}+p}{\partial p/ \partial {\cal{E}}} -\frac{6}{4\pi r^2}\right)}{r-2M} 
\nonumber \\
&& -4\left[\frac{M + 4\pi r^3 p}{r^2(1-2M/r)}\right]^2, 
\end{eqnarray}
along with the TOV equations (\ref{lambda}, \ref{nu},\ref{pres}, and \ref{mass}) with the appropriate boundary conditions, as given in \cite{tuhin}. 
After solving these equations and obtaining the values $M$, $R$, $k_2$ $etc.$, one can compute the dimensionless tidal polarizability as: $\Lambda = 2/3 k_2 C^{-5}$.


\section{Results and discussions}{\label {result}}

\subsection{Effect of dark matter on the EOS of neutron star} {\label {eos-dm}}

The main goal of this work is to investigate the possible effects of DM on the properties of NSs, which depend entirely on the nature of the EOS of the object. It should be noted here that contrary to
\cite{ellis18,ellisplb}, in the present work the DM particles are everywhere inside the NSs and so they do not form a compact core. Consequently, additional peaks in post-merger power spectral density (PSD), as studied in \cite{ellisplb}, most likely will not be produced. We have obtained the EOSs within E-RMF model using recently developed parameter sets, such as IOPB-I \cite{iopb1} and G3 \cite{G3}. 
The merit of these parameter sets is that they pass through or relatively close to the low as well as high density region \cite{iopb1}. The EOS corresponding to the G3 set is softer than the one corresponding to the IOPB-I parameter set \cite{iopb1}. For comparison reasons, we also consider the NL3 parameter set \cite{lala97}, which is one of the best known and widely used set. The EOS corresponding to the NL3 set is the stiffest among the chosen parameter sets. Table \ref{table1} shows values of the coefficients of Eqs. \ref{eq20}, \ref{eq21} for the parameters sets considered here. To obtain the EOS of an NS in the presence of DM, we need the values of the parameters for the DM part in the full Lagrangian (Eq. \ref{lag-tot}).
The values of the quantities involved in Eq. \ref{lag-tot} have already been mentioned in Sub-section \ref{EOS-NS} except 
for $M_{\chi}(\equiv M_{DM})$ and the Yukawa coupling $y$. It has been stated that the values of $y$ are obtained using Eq. \ref{sicross}, varying the mass 
of the DM particle in the range specified in Eq. \ref{mdm}. 

\begin{figure}
        \includegraphics[width=1.0\columnwidth]{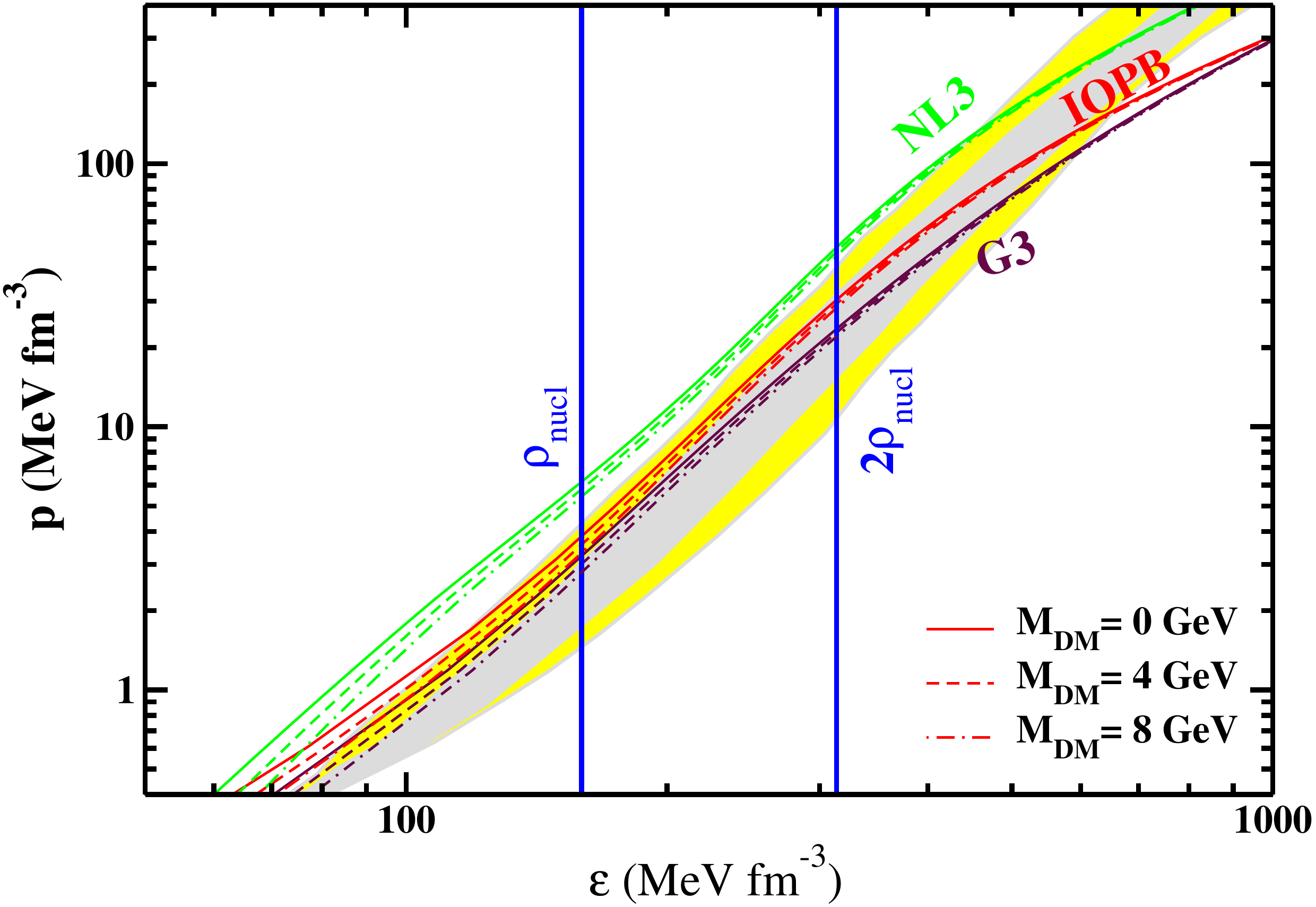}
        \caption{(color online) The EOSs of NSs in the presence of DM, corresponding to G3, IOPB-I, and NL3 parameter sets. 
The bold lines labelled as $M_{DM}=0$ GeV represent the EOSs without considering DM. The dashed and dash-dotted lines represent the 
EOSs in the presence of DM with a WIMP mass $M_{DM}=4$ GeV and  $M_{DM}=8$ GeV, respectively. The grey (yellow) 
shaded region correspond to the 50\% (90\%) posterior credible limit from the GW data \cite{abbott18}.}
        \label{eos}
\end{figure}

In Figure \ref{eos}, the EOSs of NS corresponding to the G3, IOPB-I, and NL3 parameter sets are shown for two different values of DM particle mass i.e., $M_{DM}=4$ GeV (dashed curve) and $M_{DM}=8$ GeV (dash-dotted curve). 
For comparison reasons, the EOSs without DM $M_{DM}=0$ GeV (bold curves) are also shown for all the parameter sets. 
The grey and yellow shaded regions represent the 50\% and 90\% confidence level of EOS, obtained from GW data \cite{abbott18}.
This was done using the spectral EOS parameterization with the condition that the EOS must support at least a $1.97~M_{\odot}$ star \cite{abbott18}. 
In this way, the pressure posterior band \cite{abbott18} shrinks about three times from the prior pressure \cite{abbott17} (not shown here). It is remarked in \cite{abbott18} that posterior EOS becomes softer than the prior EOS. The vertical lines (blue lines) represent the nuclear saturation density and twice its value. These densities are assumed to almost correlate with bulk macroscopic properties of NS \cite{ozel16}. The pressure at twice of the nuclear saturation density is measured to be $21.88^{+16.88}_{-10.62}$ MeV-fm$^{-3}$ \cite{abbott18}. 
We see that the presence of DM inside NS softens the EOS. A higher mass of the DM particle has a stronger impact in softening the EOS. The shift of the EOSs curves with the increase of the mass of the DM particle can be easily seen. It can be observed from the figure that G3 and IOPB-I EOSs with and without DM pass through the 90\% credible limit of the experimental band (posterior EOS) at and around the nuclear saturation density $\rho_{nucl}$. These EOSs pass through the 50\% 
confidence level too at slightly larger energy density than $\rho_{nucl}$, while at larger nuclear density, these EOSs become 
softer than the shaded band. However, NL3 EOSs satisfy the 90\% as well as 50\% confidence level of posterior EOSs only at a very large value of the energy density. 
The effects of DM on the EOSs are consistent with what was obtained in \cite{panta,tuhin}, where the authors fixed the mass of the DM particle and varied the wave number of the DM particle. In \cite{panta,tuhin}, the effects of DM were larger 
due to the fact that there the SM Higgs boson was the mediator and the mass of the DM particle was taken as 200 GeV. 
In this work, however, we have considered lighter DM particle 
and Higgs bosons, as was mentioned before, which can accommodate for scattering results consistent with the DAMA/LIBRA experiment. It is important to mention that the EOS becomes stiffer when considering the DM haloes around NS \cite{nelson}. 
In this case, an enhancement in structural properties is reported in \cite{nelson}. 

\begin{table}
\caption{The parameters involved in EOS (Eqs. \ref{eq20}, \ref{eq21}) corresponding to IOPB-I \cite{iopb1}, G3 \cite{G3}, 
and NL3 \cite{lala97} parameter sets are listed. The mass of nucleon  
$M$ is 939.0 MeV in all the sets.  All the coupling constants are dimensionless, except $k_3$ which is in fm$^{-1}$.}
\scalebox{1.5}{
\begin{tabular}{cccccccccc}
\hline
\hline
\multicolumn{1}{c}{}
&\multicolumn{1}{c}{NL3}
&\multicolumn{1}{c}{G3}
&\multicolumn{1}{c}{IOPB-I}\\
\hline
$m_{s}/M$  &  0.541  &    0.559&0.533  \\
$m_{\omega}/M$  &  0.833  &  0.832&0.833  \\
$m_{\rho}/M$  &  0.812 &   0.820&0.812  \\
$m_{\delta}/M$   & 0.0  &   1.043&0.0  \\
$g_{s}/4 \pi$  &  0.813   &  0.782 &0.827 \\
$g_{\omega}/4 \pi$  &  1.024  &   0.923&1.062 \\
$g_{\rho}/4 \pi$  &  0.712  &   0.962 &0.885  \\
$g_{\delta}/4 \pi$  &  0.0  &   0.160& 0.0 \\
$k_{3} $   &  1.465  &    2.606 &1.496 \\
$k_{4}$  &  -5.688  &  1.694 &-2.932  \\
$\zeta_{0}$  &  0.0  &  1.010  &3.103  \\
$\eta_{1}$  &  0.0  &   0.424 &0.0  \\
$\eta_{2}$  &  0.0  &   0.114 &0.0  \\
$\eta_{\rho}$  &  0.0 &  0.645& 0.0  \\
$\Lambda_{\omega}$  &  0.0 &  0.038&0.024   \\
\hline
\hline
\end{tabular}}
\label{table1}
\end{table}

\subsection{Neutron star observables}{\label {observables}}

\begin{figure}
        \includegraphics[width=1.0\columnwidth]{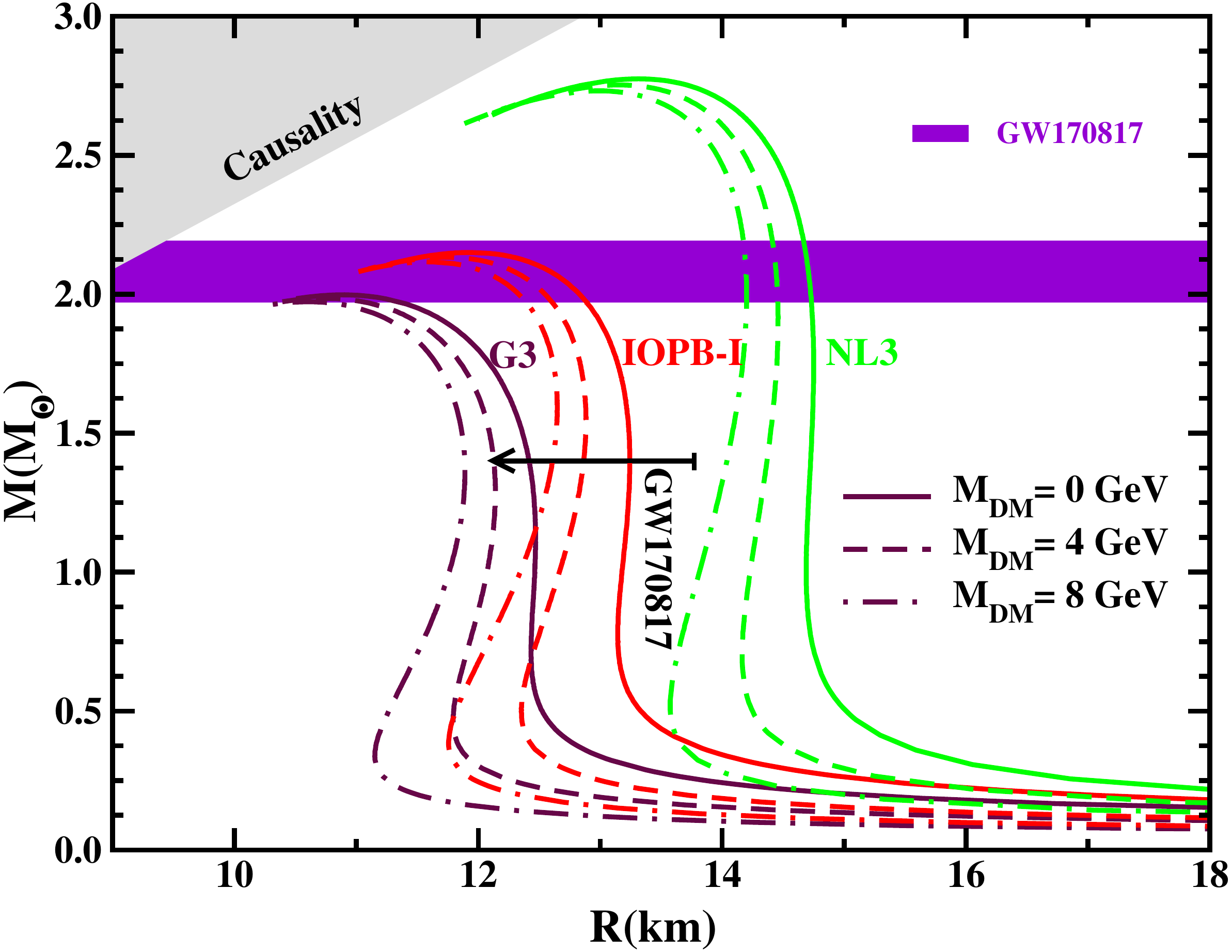}
        \caption{(color online) The mass-radius profile for NSs in the presence of DM corresponding to the 
 IOPB-I \cite{iopb1}, G3 \cite{G3}, and NL3 \cite{lala97} parameter sets. The recent constraints
on the mass \cite{rezzo18} and radii \cite{fattoyev18} of NS are also shown. The grey shaded region shows the causality region \cite{lattimer07}.}
        \label{mr}
\end{figure} 

The mass-radius profile for an NS is presented in Fig. \ref{mr} using the EOSs as shown in Fig. \ref{eos}. The violet band represents the maximum mass range for a non-rotating NS \cite{marg17,rezzo18,antoni13}. 
This band also satisfies the precisely measured mass of NS, such as PSR J0348+0432 with mass $(2.01 \pm 0.04) M_\odot$ \cite{antoni13}. 
These results imply that the theoretically predicted masses of NSs should reach the limit $\sim 2.0~M_\odot$. 
The black arrow represents the radius at the canonical mass of NS \cite{fattoyev18} with the maximum value $R_{1.4} \leq 13.76$~km. 
As anticipated, the mass-radius (MR) profiles are shifted downwards in the presence of DM inside a NS. The small effect of the DM on EOS 
produces a significant shift of the MR profile to the left with a slightly lower highest mass. The bold, dashed, and dash-dotted lines correspond to the same representation 
throughout this work, as mentioned for EOS figure (Fig. \ref{eos}). The NL3 set, being the stiffest among the considered parameter sets, predicts large mass and radius. The G3 and IOPB-I EOSs with DM predict the maximum masses of NS that satisfy 
the mass range constrained in \cite{rezzo18} from the prior GW data \cite{abbott17}, while GW170817 data rule out the NL3 EOSs. 
On considering the large value of DM wave number, the EOSs for the NL3 set can be significantly reduced to satisfy the GW170817 
mass range. In the figure, the lowering in the maximum mass of NS for the EOSs with DM is small. The effects of DM are more important for masses below the highest mass. In other words, the radius is more reduced at a fixed mass other than the maximum mass. 
Among the three parameter sets considered here, the IOPB-I EOSs with and without DM satisfy the radius range at canonical mass constrained by the event GW170817 \cite{fattoyev18}.

\begin{figure}
        \includegraphics[width=1.0\columnwidth]{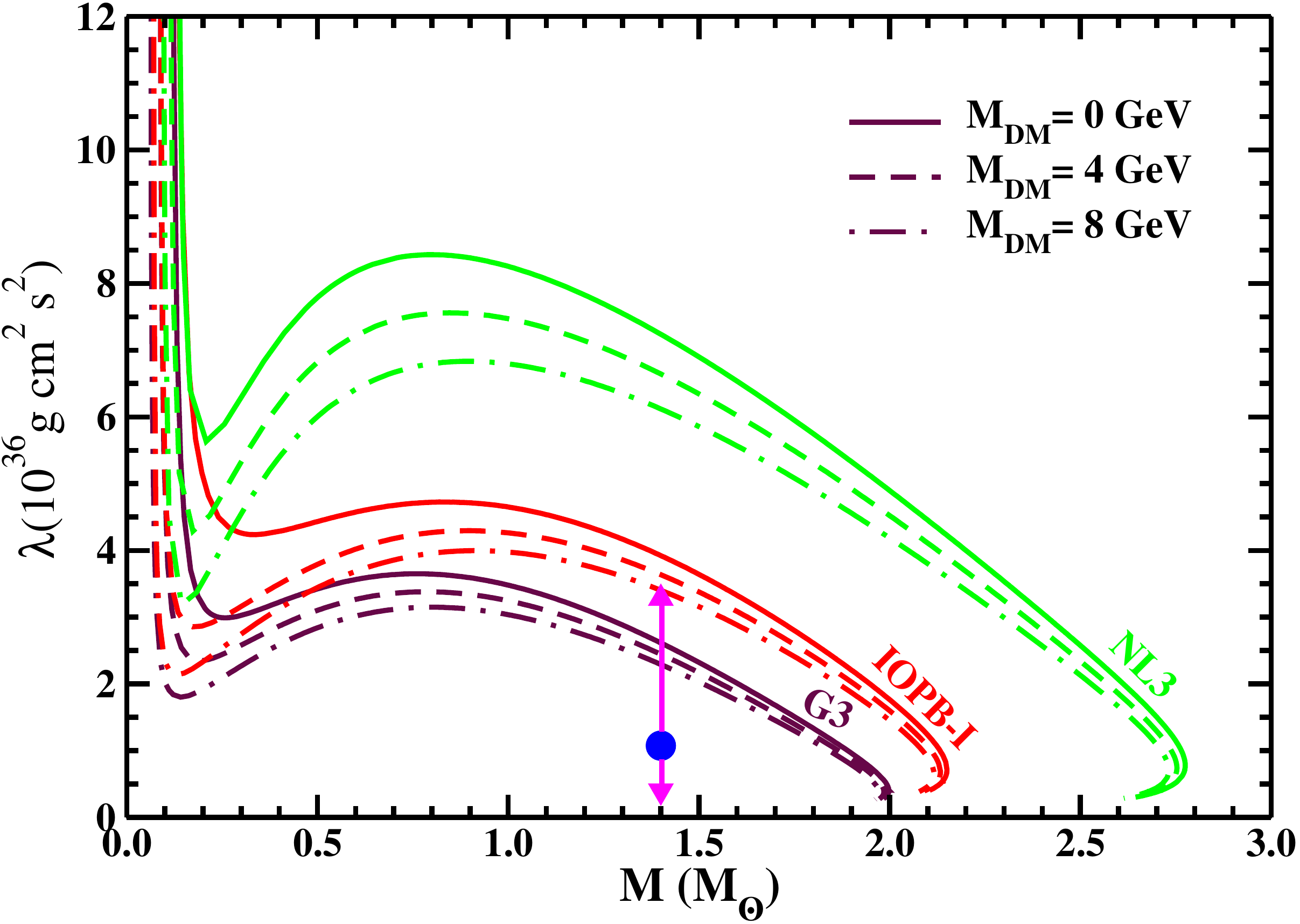}
        \caption{(color online) Tidal deformability $\lambda_2$ as a function of NS mass corresponding to the IOPB-I, G3, 
and NL3 EOSs. The dashed and dash-dotted lines represent the EOSs in the presence of DM with the neutralino mass  
$M_{DM}=4$ GeV and  $M_{DM}=8$ GeV, respectively. The blue circle with the arrow bar represent the $\lambda_2$ value 
at 1.4M$_\odot$ NS mass obtained from GW data \cite{abbott18}.}
        \label{lm}
\end{figure} 

\begin{figure}
        \includegraphics[width=1.0\columnwidth]{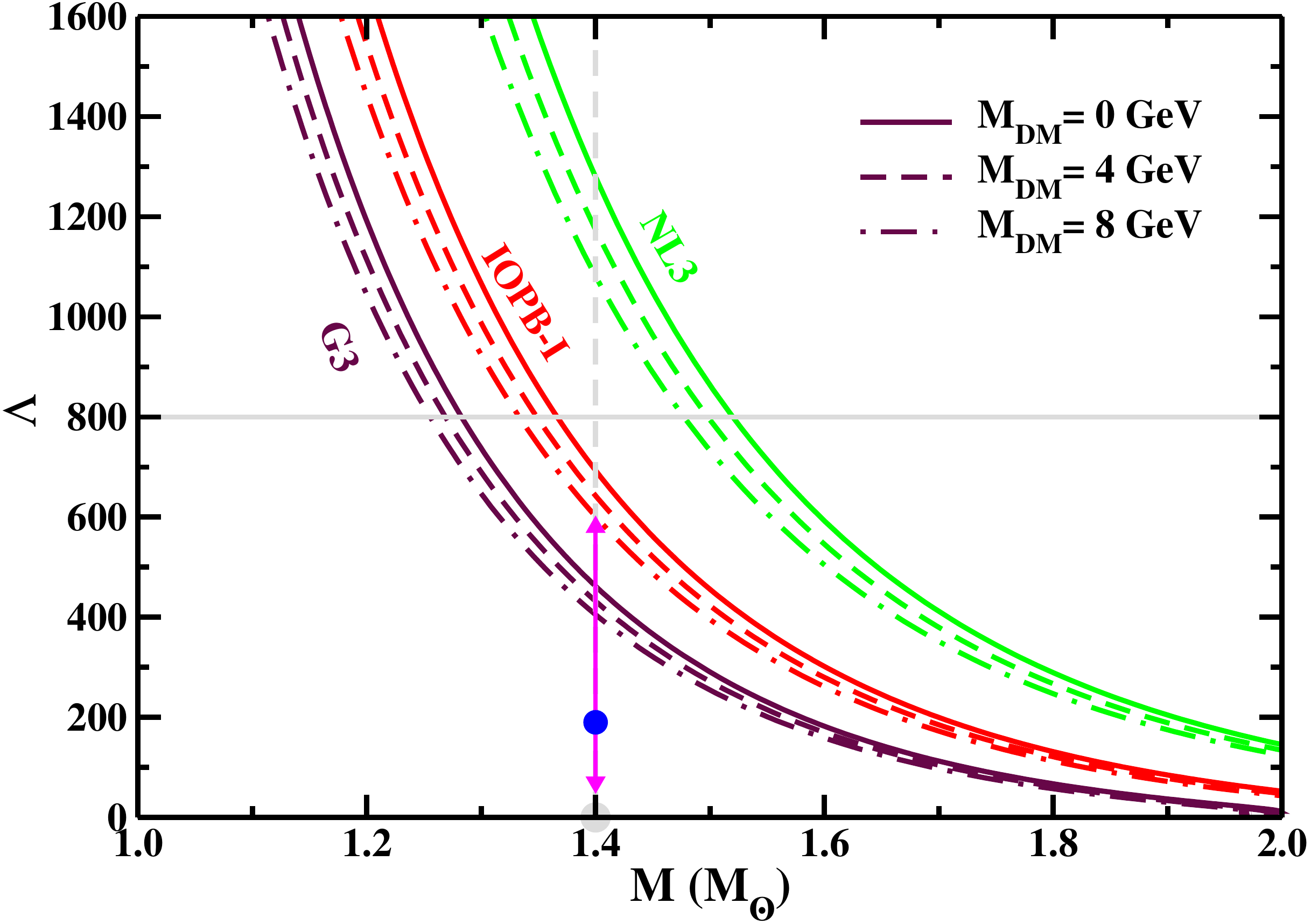}
        \caption{(color online) The dimensionless tidal deformability parameter ($\Lambda$) as a function of NS mass 
corresponding to the IOPB-I, G3, and NL3 EOSs. The dashed grey line represent the canonical mass of NS. 
However, the bold grey line shows the upper limit of $\Lambda$ value from GW170817 data \cite{abbott17}. 
The blue circle with the error bar represent the $\Lambda_{1.4}$ value for posterior GW170817 data \cite{abbott18}. }
        \label{LM2}
\end{figure}

The tidal deformability of NS depends on its mass quadrupole, which is developed due to the tidal gravitational field of another component of NS binaries, as discussed above. It quantifies mainly the surface part of NS. 
We have calculated the tidal polarizability for $l=2$ perturbation, i.e., $\lambda_2$. Recently, tidal deformability was discussed 
for the GW170817 data \cite{abbott17}. It is clear from its definition (Eq. \ref{lam2}) that $\lambda_2$ depends on the radius of a star and on its tidal love number $k_2$, which describes the internal structure of NS. 
As the radius of NS increases, $\lambda_2$ values grow and the surface becomes more deform. It simply means that soft EOSs predict 
less value for $\lambda_2$. In Figure \ref{lm}, we plot $\lambda_2$ for the chosen EOSs with and without DM. The blue circle with the arrow bar (error bar) represents the $\lambda_2$ of an NS at the mass 1.4M$_\odot$ corresponding 
to the $\hat{\Lambda}_{1.4} = 190^{+390}_{-120}$, which is constrained from the GW data \cite{abbott18} at 90\% confidence level. 
The NL3 set predicts large values for the tidal deformability and hence large deformation. The $\lambda_2$ corresponding to the NL3 EOSs, 
even in the presence of DM, does not pass through the experimental range at the canonical mass. 
On the other hand, $\lambda_2$ curves for the G3 EOSs lie within the observationally allowed region. However, 
the IOPB-I EOS at neutralino mass $M_{DM}=8$~GeV predicts a $\lambda_2$ value that just satisfies the upper range of 
the experimental $\lambda_2$ value. The shift of the curves in the presence of DM can easily be noticed from the figure. The significant changes in $\lambda_2$ due to the DM occur at the canonical mass of NS. We also show the dimensionless tidal deformability ($\Lambda$) of a single NS in Figure \ref{LM2}. The blue circle represents $\Lambda$ at the canonical mass of NS from GW170817 posterior data \cite{abbott18}, the numerical value of which is given above.

\begin{figure}
        \includegraphics[width=1.0\columnwidth]{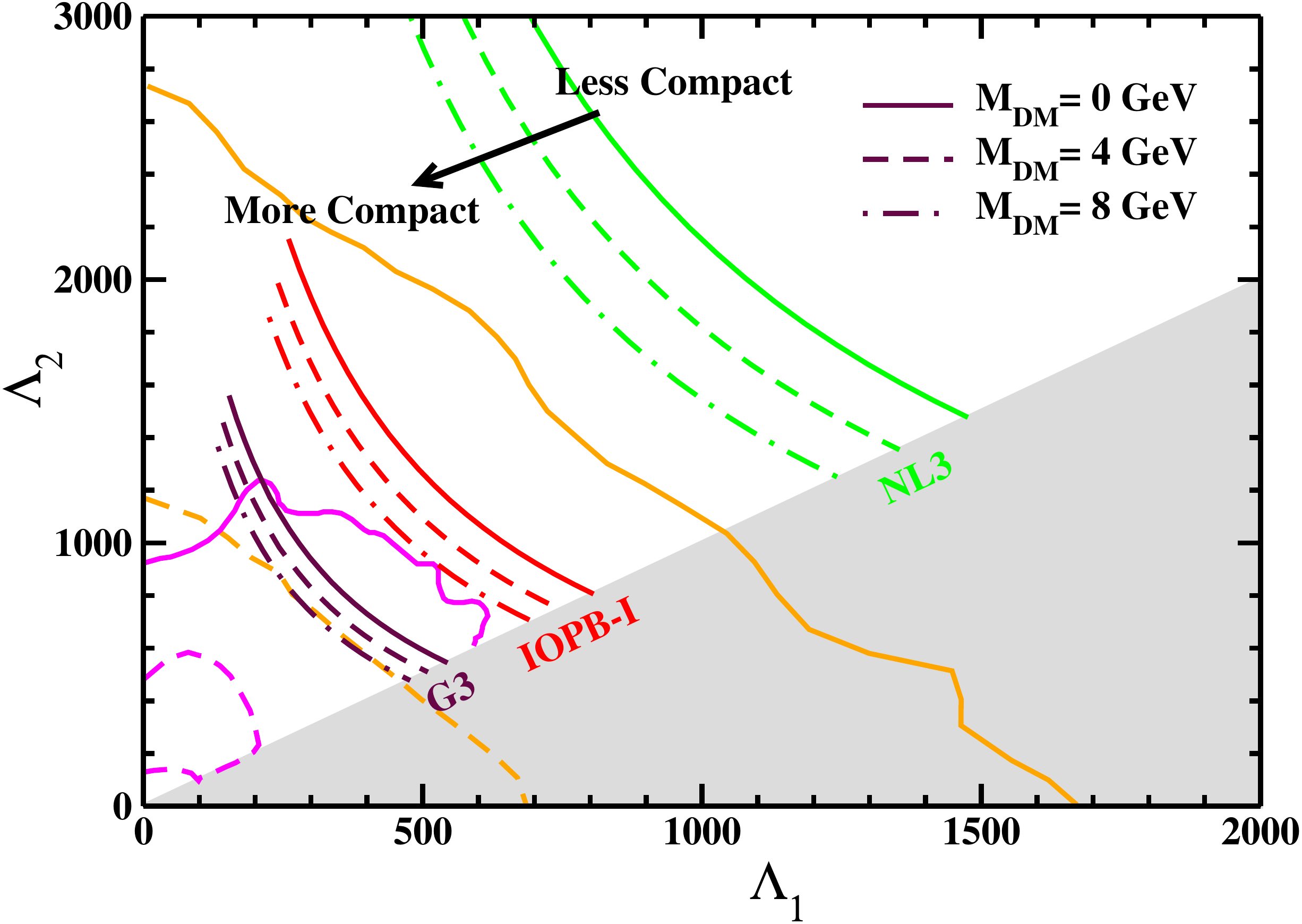}
        \caption{(color online) Dimensionless tidal deformability $\Lambda$ generated by using IOPB-I, G3, and NL3 EOSs with DM and
without DM inside the NS. The calculated values are compared with 50\% (dashed) as well as 90\% (bold) probability contour for 
the case of low spin, $|\chi| \leq 0.05$, and represented by Orange and Magenta color for, respectively, prior \cite{abbott17} 
and posterior \cite{abbott18} GW170817 data. }
        \label{tidal}
\end{figure}

In Fig. \ref{tidal} we display the dimensionless tidal deformabilities $\Lambda_1 $ and $\Lambda_2 $ of a binary NS 
corresponding to the G3, IOPB-I, and NL3 parameter sets. Here, we consider the tidal deformability constraint from 
GW170817 observation on EOSs of NS in the presence of DM. The individual dimensionless tidal deformabilities 
$\Lambda_1 $ and $\Lambda_2 $ correspond to high mass $m_1$ and low mass $m_2$ of BNS. We vary the mass $m_1$ 
in the range $1.365 < m_1/M_\odot < 1.60$, and determine the range of $m_2$ by fixing the chirp mass as ${\cal{M}}_c = 1.188~M_{\odot}$.
It can be seen in the figure that the G3 and IOPB-I sets are in excellent agreement with the $90\%$ (bold line) 
probability contour of prior GW170817, shown by orange curves \cite{abbott17}.
We also show the recently re-analyzed results of GW170817 data in magenta color \cite{abbott18}. The figure shows that only the curves 
corresponding to the G3 EOS with and without DM lies within the 90\% confidence level allowed region of prior as well as 
posterior GW170817 data.
The shaded part (grey color) in the figure marks the $\Lambda_2 < \Lambda_1$ region that is naturally excluded for a common realistic EOS \cite{abbott18}. The analysis of \cite{abbott18} suggests that soft EOSs, which predict lower values for $\Lambda$ are favored over stiffer EOSs. 
For the EOSs corresponding to DM admixed NSs, the curves are shifted to the left and predict lower values for $\hat{\Lambda}$ corresponding to less compact NSs. The NL3 EOSs lie outside the 90\% confidence lever region (bold line) of prior (orange) 
as well as posterior (magenta) analysis. 
On adjusting the parameters of the DM Lagrangian for an EOS, the curve can be shifted even more to the left. That way, the parameters of the DM Lagrangian can be optimized satisfying the GW170817 constraints.

\begin{figure}
        \includegraphics[width=1.0\columnwidth]{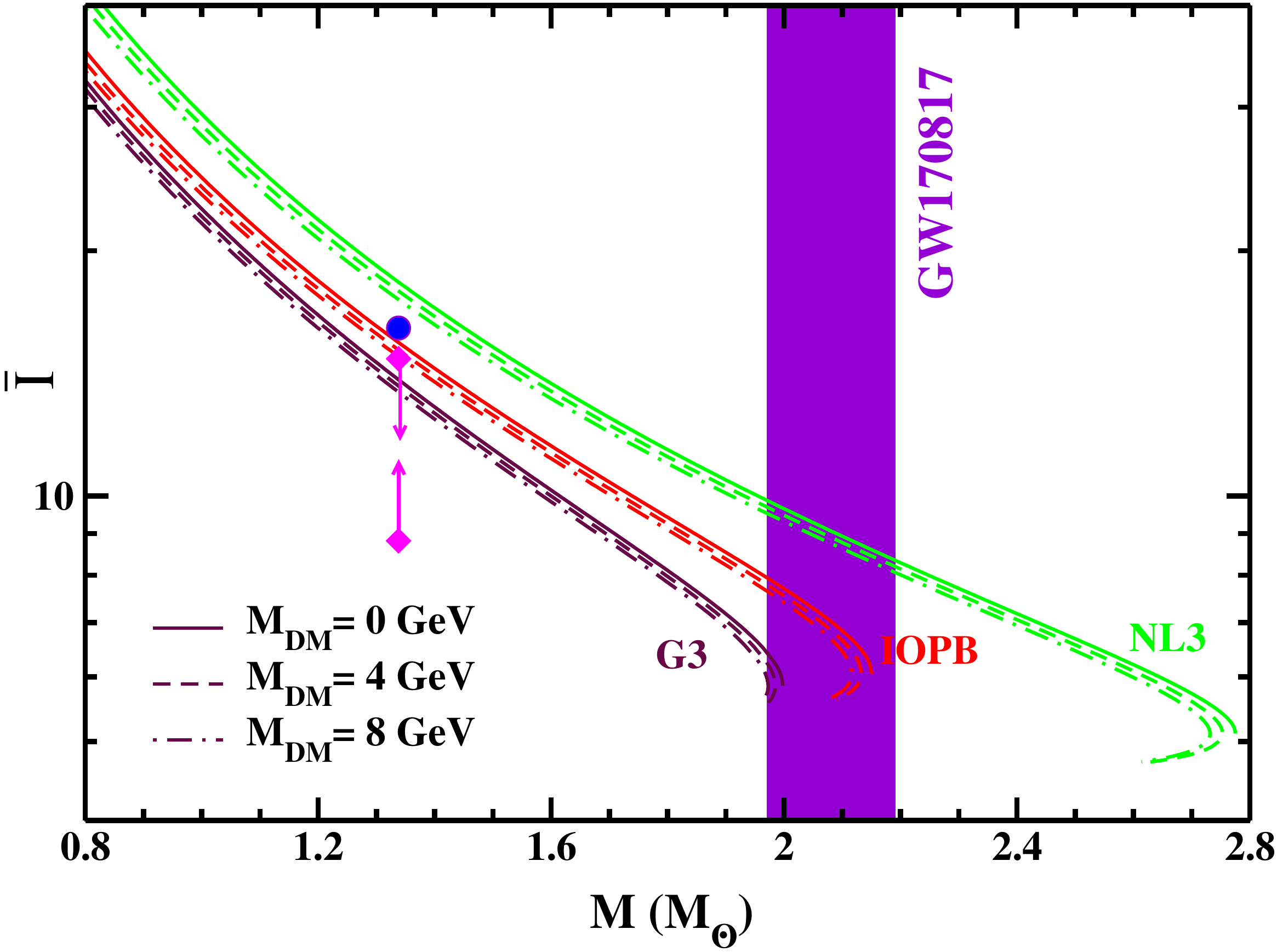}
        \caption{(color online) The dimensionless moment of inertia $\bar I$ as a function of NS mass for EOSs shown in Fig. \ref{eos}. 
The overlaid arrows represent the constraints on MI of PSR J0737-3039A set \cite{bharatAPJ} 
from the analysis of GW170817 data \cite{abbott18}. 
The circle shows the upper bound from the minimal-assumption analysis of Ref. \cite{abbott17}. 
From reference point of view, the mass range of NS, constrained from GW data \cite{rezzo18}, is also shown by the violet band.} 
        \label{moi}
\end{figure}

The moment of inertia (MI) of NSs strongly depends on the structure of the object. For a slowly rotating NS the moment of inertia is computed solving Eqs. \ref{mi}, \ref{bound.}, and \ref{unitlessw.} together with the TOV equations. It is one of the most important macroscopic quantities that can be used to constrain the EOS of NSs. The MI of the binary pulsar, PSR J0737-3039, is expected to be determined within $\sim 10\%$ accuracy by measuring its angular momentum \cite{lyne04,kramer09,lattimer05}. The mass distribution of a NS, the final stage of the BNS merger, and $r-$ process nucleosynthesis are determined by the EOS, and therefore a precise measurement of the MI, tidal deformability, etc are very important. In Fig. \ref{moi} we plot the dimensionless MI, which decreases with the mass of an NS. Stiffer EOSs predict a larger MI for a given mass of a NS. In the presence of DM, the soft nature of EOSs generates a lower MI. The overlaid arrows in the figure indicate the MI of PSR J0737-3039, constrained by the analysis of GW170817 \cite{abbott18}, while the circle represents the upper bound on MI from minimal assumption analysis \cite{abbott17}.

\section{Summary and Conclusions}{\label{summary}}

In summary, we have analyzed the effect of DM on the properties of NSs. 
We have considered a generic WIMP of fermionic nature (one may have in mind the lightest neutralino), which are trapped inside the object. The WIMPs interact with baryonic matter through the exchange of light Higgs bosons. We have adopted the $\sim 10~GeV$ WIMP hypothesis, as suggested by the DAMA/LIBRA results. As it has been shown in the literature that such a scenario can be realized within NMSSM, one can have in mind the lightest CP-even eigenstate of the NMSSM as the mediator Higgs boson. The EOSs of NSs are generated using the E-RMF Lagrangian density including the interaction Lagrangian density of DM with baryonic matter, and applying the $\beta-$ equilibrium and charge neutrality conditions. 
Within E-RMF, we have used the recent parameter sets, such as G3 and IOPB-I, along with the older and widely accepted NL3 set. Out of the three EOSs considered here, G3 is the softest one and predicts relatively small values of NS observables in agreement 
with the GW170817 results. We have observed that the presence of DM in NS softens the EOS, which results in lowering the values of NS observables, such as mass, radius, tidal deformability, and even moment of inertia. We have imposed the constraints 
from GW170807 on the mass, $\lambda_2$ values, dimensionless tidal deformability $\Lambda$ and MI of NSs. The effects of DM are small at the maximum values of NS mass, while its impact is more significant in the mass region other than the maximum mass.


{\bf Acknowledgments:}

It is a pleasure to thank F.~Cappella for enlightening discussions and correspondence. We are thankful to the referees 
for enhancing the quality of the paper.  
A.~Q. would like to acknowledge DST-INSPIRE for providing financial support in the form of fellowship with order number 
No./DST/INSPIRE Fellowship/2016/IF160131, and Institute of Physics, Bhubaneswar, for providing the hospitality during the work. 
G.~P. thanks the Fun\-da\c c\~ao para a Ci\^encia e Tecnologia (FCT), Portugal, for the financial support to the Center for Astrophysics and Gravitation-CENTRA, Instituto Superior T\'ecnico, Universidade de Lisboa, through the Grant No. UID/FIS/ 00099/2013.
B.~K. thanks the Navajbai Ratan Tata Trust for providing partial support for this work. This work is supported in part by JSPS KAKENHI Grant No.18H01209.


\end{document}